\documentclass[preprint,review,12pt,leqno]{elsarticle}
\usepackage{float}
\usepackage{graphicx}
\usepackage{grffile}
\usepackage{multirow}
\usepackage{gensymb}
\usepackage{booktabs}
\usepackage{tabularx}
\usepackage{footnote}
\usepackage{amsmath} 
\usepackage{color,amsmath,amssymb}
\usepackage{color}
\usepackage{comment}
\usepackage{amssymb}
\usepackage{subcaption}
\usepackage{tikz}
\usepackage{braket}

\begin{document}

\title{Determination of approximate quantum labels based on projections of the total angular momentum on the molecule-fixed axis}

\author{Eamon K. Conway}
 \address{Center for Astrophysics $|$ Harvard \& Smithsonian,  Atomic and Molecular Physics Division, Cambridge, MA, USA. 02138}
\address{Department of Physics and Astronomy, University College London, Gower Street, London WC1E 6BT, United Kingdom}

\author{Iouli E. Gordon}
 \address{Center for Astrophysics $|$ Harvard \& Smithsonian,  Atomic and Molecular Physics Division, Cambridge, MA, USA. 02138}

\author{Oleg L. Polyansky}
 \address{Department of Physics and Astronomy, University College London, Gower Street, London WC1E 6BT, United Kingdom} 
\address{Institute of Applied Physics, Russian Academy of Sciences, 46 Ulyanov Street, Nizhny Novgorod, 603950, Russia}
 \author{Jonathan Tennyson}
 \address{Department of Physics and Astronomy, University College London, Gower Street, London WC1E 6BT, United Kingdom}

\begin{abstract}

Molecular line lists, particularly those computed for high temperature applications, often have very few states assigned local quantum numbers. These are often important components for accurately determining line shape parameters required for radiative transfer simulations. The projection of the total angular momentum onto the molecule fixed axis ($k$) is investigated in the Radau internal coordinate system to determine when it can be considered a good quantum number. In such a coordinate system, when the square of the $k^{th}$ component of the wavefunction is greater than one half, then we can classify $k$ as a good quantum number in accordance with the theorem of Hose and Taylor. Furthermore, it is demonstrated that when this holds true, oblate and prolate quantum labels $K_{a}$ and $K_{c}$ can reliably be predicted. This is demonstrated for the water and ozone molecules. 

\end{abstract}

\maketitle

\section{Motivation}

High resolution molecular spectroscopy routinely labels transitions and hence energy levels with quantum numbers which specify the (sometime approximate) constants of
motion for the system. These labels are a mixture of rigorous (symmetry) constants of motion and more approximate labels which are usually derived from the underlying
model. These quantum labels are important for inter-comparison, identification of states, physical understanding, obtaining
both rigorous and approximate selection rules and a variety of other uses some of which are discussed below.

For small molecules, spectroscopic data is increasingly being generated using variational nuclear motion programs. While these programs in general use the rigorous
quantum numbers to simplify the solution of the nuclear motion Schr\"{o}dinger equation, there is no guarantee that the zeroth order model used by these programs will yield
the approximate quantum numbers generally used.
The DVR3D \cite{DVR3D} suite of programs use variational methods  to solves the exact nuclear-motion Schr\"{o}dinger equation within the Born-Oppenheimer approximation 
for triatomic molecules. DVR3D has been used used in the calculation of numerous line lists, some of the most notable are H$_{2}$S \cite{jt640}, CO$_{2}$ \cite{jt625}, HCN \cite{jt689}, SO$_{2}$ \cite{jt635}, H$_{3}$$^{+}$ \cite{jt748} and of course H$_{2}$O \cite{jt734}. The ExoMol \cite{jt631,jt810} project has seen numerous high temperature line lists \cite{jt734,jt804,20ChTeYu} developed for use in applications where high temperatures are expected, such as exoplanets, cool stars and combustion experiments. 

In DVR3D the molecules with  like atoms can be treated in the C$_{\text{2v}}$ symmetry representation and four states are possible: A$_{\text{1}}$, A$_{\text{2}}$, B$_{\text{1}}$ and B$_{\text{2}}$. Vibrational symmetry and the  parity quantum labels are selectively chosen based on the eigenstate of interest. The DVR3D setup only provides these rigorous quantum labels such as the total angular momentum ($J$), rotationless parity ($p$) and interchange symmetry ($q$). 

In general, it is useful to assign states with more generic/local labels, such as for asymmetric top species the rotational quantum labels $K_{a}$ and $K_{c}$, along with vibrational quanta $\nu_{1}$, $\nu_{2}$ and $\nu_{3}$ symbolizing symmetric stretch, bend and asymmetric stretch, respectively. These quantum labels become particularly important in the determination of line shape parameters, which can be a function of the vibrational quanta that are exchanged in a transition \cite{REGALIA2019126,Vispoel2019}, as well as being dependent on the change in rotational quanta. Indeed it is well established
that it is the rotational quantum numbers $(J,K_a,K_c)$, which are most important for characterizing line broadening parameters \cite{jt483,11MaTiLa}. Quantum labels can be used to transfer information
between different isotopologues \cite{jt665,jt817}, providing the symmetry is conserved on isotopic substitution.
These labels  can also serve as unique fingerprints in line lists which can contain thousands if not millions or billions of transitions that we often see in high temperature line lists. In such high temperature line lists that have an enormous quantity of transitions, numerous states do not have any local quantum labels and hence lack important spectroscopic information. While rigorous dipole selection rules depend on the total angular momentum $J$ and parity considerations, there are strong
propensity rules which usually  favour small changes in $K_{a}$, making assigning values to this number particularly important.

\v{S}mydke and Cs\'{a}sz\'{a}r \cite{19SmCsXX.H2O} showed that vibrational labels could be automatically assigned to variationally calculated states by computing reduced-density matrices. The procedure was shown to be highly reliable up to approximately 25 000 cm$^{-1}$. Other variational programs/methods \cite{VTET,TROVE} that solve the Schr\"{o}dinger equation make use of different basis sets that can be used to assign an approximate label on the degree of vibrational excitation \cite{05YuThPa.PH3,97PaScxx.H2O}. Other methods have also been proposed for providing the approximate vibrational
labels \cite{07JuTa,10MaFaSz,12SzFaCs,20MaGu}.
There has, however, been  little work performed on assigning  rotational quantum numbers from variational wavefunctions. However,it should be recognised that for molecules which go from bent to linear it has been demonstrated \cite{jt234} that it is not possible to characterize all states using a single set of quantum numbers. 

MARVEL (Measured Active Rotation Vibration Energy Levels) \cite{jt412} is an algorithmic procedure where high-quality spectroscopic information on a particular molecule is analyzed in a spectroscopic network \cite{ 11CsFuxx.marvel} with the aim of creating an accurate set of empirical energy levels. Variational line lists can often be complete, or at least very extensive, thereby possessing many, if not all energy levels for a molecule, although transition frequencies are less precise than their experimental counterparts. For those well studied molecules such as water, where numerous experiments have observed a single transition, it can become very difficult determining what is the preferred set of data. In such cases, the MARVEL algorithm can use all known information of the molecule to determine the best set of data. This procedure has been hugely successful for many molecules \cite{jt809,jt791,jt764,jt740,jt718,jt705,jt672,jt637,jt608,11FaMaFu.marvel,13FuSzFa.marvel,13FuSzMa.marvel,19FuHoKoSo} and for some, can determine energy levels to kHz precision \cite{20ToFuSiCs}. For tri-atomic systems (or larger), the MARVEL approach is currently limited to those molecules where the knowledge of local quantum labels is extensive. If energy levels could reliably be labelled with local quantum labels in variational calculations, the MARVEL algorithm could perhaps be adapted to include such results.  Similar considerations effect the use of effective Hamiltonians, which are based on expansions in local quantum numbers, and are
increasingly being used in combination with variational calculations, see Refs. \cite{jt625,jt826}.

HITRAN \cite{jt691s}, HITEMP \cite{HITEMP2010}, GEISA \cite{GEISA} and ExoMol \cite{jt810} are examples of spectroscopic databases widely used in remote sensing, atmospheric modeling and exoplanetary research. The line lists are comprised of experimental measurements, variational calculations and semi-empirical models such as effective-Hamiltonians. Variational calculations do not rely on experimental results, although they can be used to improve accuracy \cite{jt803,jt734,11BuPoZo.H2O}, while the semi-empirical data are correlated with the pre-existence of experimental data (to model/fit). As such, a large portion of data can be variational in nature and not labelled with local quantum numbers. The existence of local quantum numbers can improve the accuracy of line shape parameters and thus the global accuracy of the line lists that possess variational data.


The Radau coordinate system is one such option that is available in DVR3D and it has been shown to be an excellent choice, where the value of $k$, the projection of $J$ on the molecule-fixed $z$-axis is found to be better than the value of $k$ obtained in the Eckart embedding \cite{20SaPoSz.H2O}. It remains to be seen as to whether this result can be further exploited to determine if it can be used to approximate local quantum labels. 

The Hose-Taylor Theorem \cite{83HoTaxx.H2O} gives the condition on when two states can be uniquely assigned quantum labels and is dependent on the overlap of the states' wavefunctions.  In this work, we will investigate the wavefunctions of the DVR3D program and apply the results of the Hose-Taylor Theorem to determine what unique quantum label(s) can be assigned when it holds true.


\section{Methods}

The Hose-Taylor Theorem \cite{83HoTaxx.H2O} states that if we have two quantum states $\Psi$ and $\Phi$, and if $\Psi$ and $\Phi$ are normalized and if the following inequality holds true
\begin{equation}\label{eqn:hose}
|\langle \Phi \ket{\Psi}|^{2} > \frac{1}{2} 
\end{equation}
then there is a one-to-one transformation between the two states, i.e we can uniquely assign local quantum numbers. Despite being such an important result, it has been little used in the field of
molecular spectroscopy. Fortunately, the wavefunctions (basis functions) used in DVR3D are indeed normalized and hence, they satisfy the normalization condition required by the theorem. 

The DVR3D \cite{DVR3D} program solves the Schr\"{o}dinger equation using a discrete variable representation (DVR) approach in three dimensions. The wavefunctions in the radial direction (R$_{1}$,R$_{2}$) are described by associated-Laguerre polynomials, while associated-Legendre polynomials are used in the angular coordinate ($\theta$). Within the DVR, we will use $\alpha$, $\beta$ and $\gamma$ to represent the quadrature points in R$_{1}$, R$_{2}$ and $\theta$. We will also consider Radau coordinates, where it has been shown that the value of $k$, the projection of $J$, the total angular momentum, onto the body fixed axis, in this system is a better choice of $k$ when compared to the traditional Eckart value \cite{20SaPoSz.H2O}. The process of developing the body-fixed Hamiltonian in a Radau coordinate system is described thoroughly in references \cite{jt10,jt14,jt45,jt56,jt66,jt96}. 

In our Radau coordinate system, the wavefunction can be written as 
\begin{equation}\label{eqn:2}
\Psi = \sum_{k=p}^{J} \sum_{\alpha\beta\gamma} \Psi ^{J,k} _{\alpha\beta\gamma}
\end{equation}
where $p$  describes the rotationless parity  of the system and takes values zero (e state) or one (f state); the total parity is given by $(-1)^{(J+1)}$. Another rigorous quantum number is the symmetry of the wavefunction with respect to interchange of identical nuclei, described by $q$; of course
 this label is only valid for molecular systems consisting of two identical atoms having the form AB$_{2}$. 

In terms of conventional asymmertric top quantum numbers, the  parity of the wavefunction can be given by $p=(-1)^{J+K_{a}+K_{c}}$  \cite{jt275}.  Therefore, a particular value of $K_{a}$ (or equivalently $K_{c}$) we can determine the value for $K_{c}$ (or equivalently $K_{a}$) if we know the rotationless parity $p$. Also, the label $q$ can also be deduced from the expression $(-1)^{\nu_{3}+K_{a}+K_{c}}$ where $\nu_{3}$ is the quanta of asymmetric stretch.

If we hypothesize that, for a single value (per energy level) of $k \in (p,p+1,...,J)$, this chosen $k$ can be considered a good quantum number if its wavefunction satisfies the Hose-Taylor theorem. In other words, if the following condition is true
\begin{equation}\label{eqn:3}
\sum_{\alpha\beta\gamma} (\Psi ^{J,k} _{\alpha\beta\gamma})^{2}   > \frac{1}{2} 
\end{equation}
then we can uniquely assign $k$ a local quantum number.  

The objective of this study is to prove this statement by determining what local quantum number $k$ represents (if any) when the Hose-Taylor theorem is satisfied. We will investigate this theory for two important, non-linear gaseous triatomic molecules: water and ozone.

For water, states will be calculated based on our recently published global potential
energy surface (PES) \cite{jt803}. The upper energy threshold is dissociation (41 145.94 cm$^{-1}$) \cite{06MaRiBo.H2O} and we consider states with a value of the total angular momentum less than or equal to 30. In total, this gives 806 317 states to analyze.

For ozone, we will consider the PES from Polyansky \textit{et al}. \cite{18PoZoMi.O3}. For these calculations, we consider a $J=50$ upper bound and set the upper limit of the energies to the dissociation threshold of 8563.5 cm$^{-1}$. Using these thresholds, we are left with 192 812 states. 

\section{Results}

Considering the water molecule first, for every state possessing an energy below dissociation, we calculate the value of $k$ that gives the largest value of $\sum_{\alpha\beta\gamma} (\Psi ^{J,k}_{\alpha\beta\gamma})^{2}$. Of the 806 317 states, we are only interested in those that satisfy Eq.~(\ref{eqn:3}). In total, there are 68 985 of these, which represents almost 10\% of the total number of states. For these, the remaining task is to determine what $k$ physically corresponds to. Without too much investigation, it becomes obvious that $k$ is a good approximate to $K_{a}$ (discussed below). To obtain the respective value of $K_{c}$, it can be determined from Eq.~(\ref{eqn:6:1}). This is possible as there are always two choices of $K_{c}$ for every $K_{a}$ (with the exception of $J=K_{a}=0$) and we know the value of $p$ and $J$. 
\begin{equation}\label{eqn:6:1}
p=(-1)^{J+K_{a}+K_{c}} 
\end{equation}

\begin{figure}
    \begin{subfigure}[]{0.5\textwidth}
    \centering
    \includegraphics[width=1.0\linewidth]{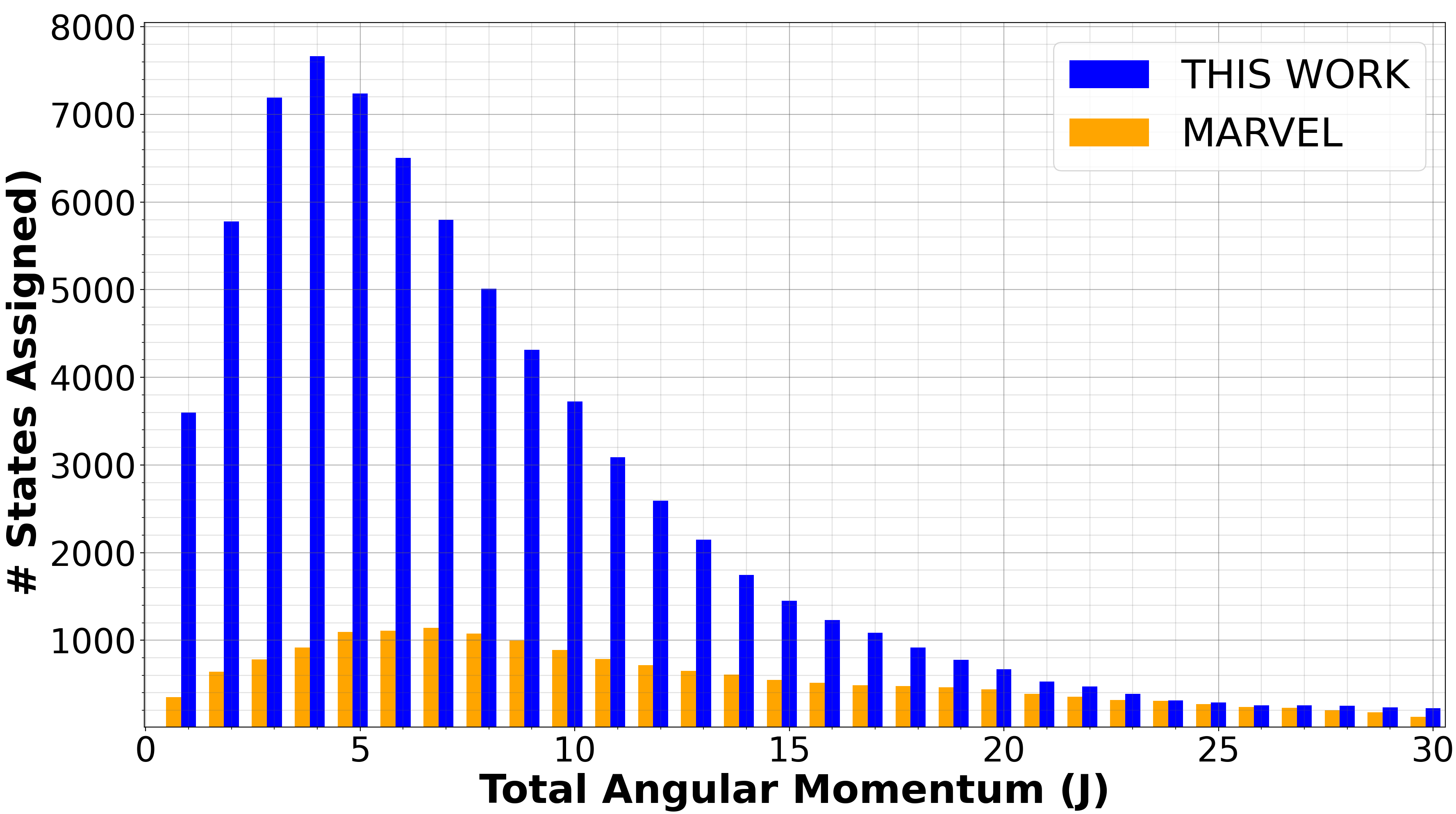}
    \caption{}
    \end{subfigure}
~
    \begin{subfigure}[]{0.5\textwidth}
    \centering
    \includegraphics[width=1.0\linewidth]{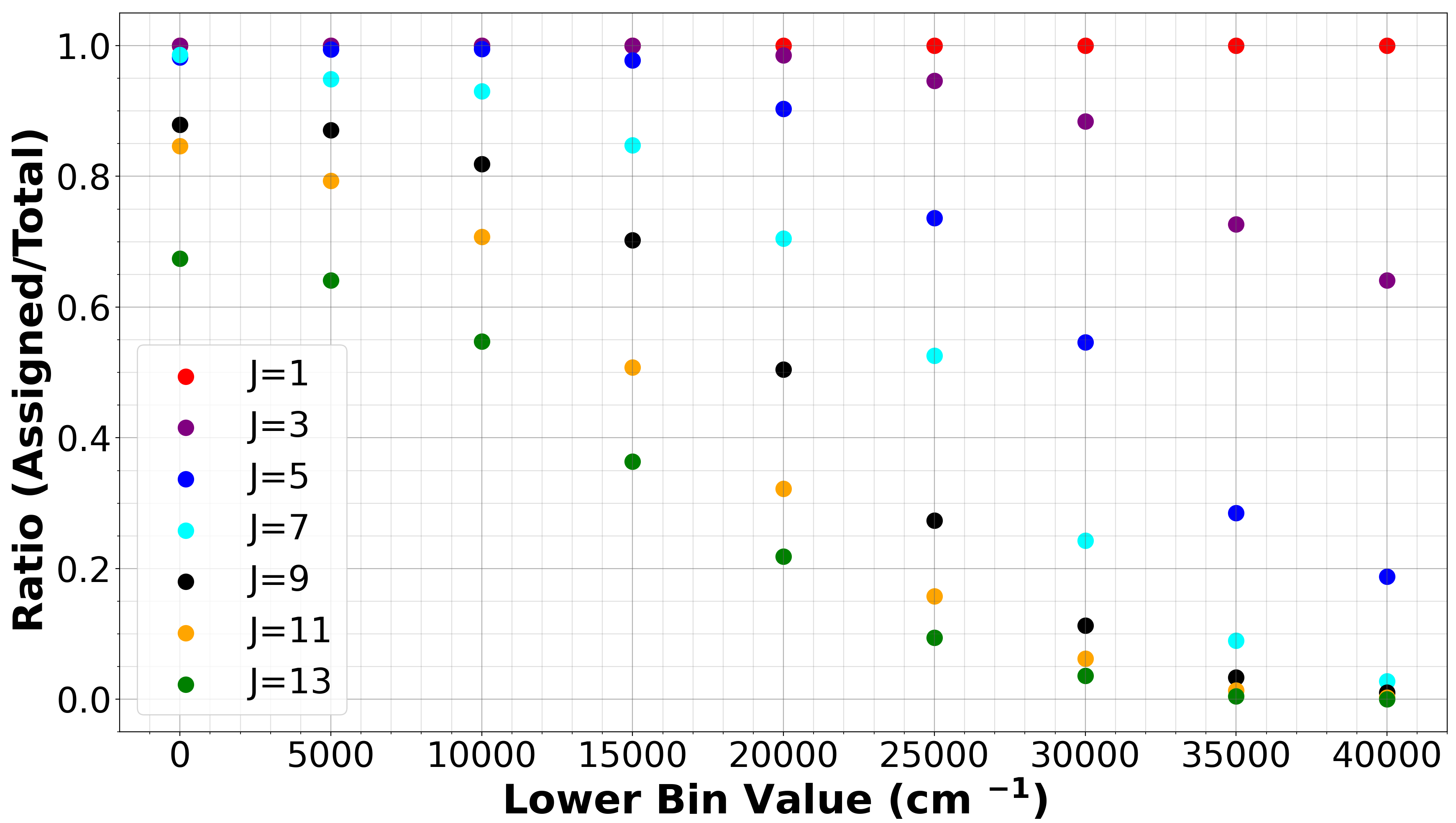}
    \caption{}
    \end{subfigure}
\caption{(a) The ratio of assigned states to the total number of available states (per $J$) as a function of increasing energies. States are grouped into bins of 5 000 cm$^{-1}$. The lower bounds of the bins are plotted. (b) The number of assigned states per $J$ value compared to what is available by MARVEL \cite{jt817}.}
\label{fig:1}
\end{figure}

In Figure~\ref{fig:1}(a), the number of possible assignments made per value of $J$ against what is provided in MARVEL \cite{jt817} (for the main water isotopologue) is plotted. The differences are quite substantial. The latest edition of MARVEL \cite{jt817} has 19 200 states available and as already explained, these are determined from experimentally measured transitions frequencies. All of these 19 200 states have local quantum assignments. Here, we manage to assign 68 985 calculated energies with accurate and reliable values of $(K_{a},K_{c})$. This is almost four times the number of assigned states available in MARVEL. For $J=4$, we manage to label 7666 energy levels with values of $K_{a}$ and $K_{c}$, while for the same $J$, MARVEL has approximately 919 levels, representing a factor seven increase. One needs to remember, however, that the MARVEL levels also have vibrational quanta associated with them, while the theoretical states do not.

It is interesting to investigate the distribution of the assignments with respect to energy. To do this, bins are created in increments of 5000 cm$^{-1}$ and the number of possible assignments are placed in the respective bin, see Table~\ref{table:1}. In Figure~\ref{fig:1}(b), the distribution of assignments are plotted for several different $J$ values as a function of the bins (lower value of each bin is plotted). As energy and/or $J$ increase, it becomes less probable to assign a value of $K_{a}$. For increasingly larger values of the total angular momentum, the number of states assignable via this method gradually reduces.

\begin{table}[H]
	\caption{Dependence of state energy on the distribution of assignments made possible for water by the Hose-Taylor theorem \cite{83HoTaxx.H2O} for $J\le15$.}
	\begin{center}
		\begin{tabular}{rrr}	
			Bin Range & Total \# States  & Assigned States \\
			\hline
			0-5000     & 745   &  652  \\
			5000-10 000     &  2963 & 2431    \\
			10 000-15 000     & 7087  & 5291    \\
			15 000-20 000     &13 334   & 8211    \\
			20 000-25 000     &  22 340 & 10 649    \\
			25 000-30 000     &   34 728&  11 372   \\
			30 000-35 000     &   51 947&  10 725   \\
			35 000-40 000     &   76 502&  9129   \\
			40 000-41 145     &   37817& 2940    \\
						\hline
		\end{tabular}
	\end{center}
	\label{table:1}
\end{table}

\begin{figure}[H]
	\includegraphics[width=1.0\linewidth]{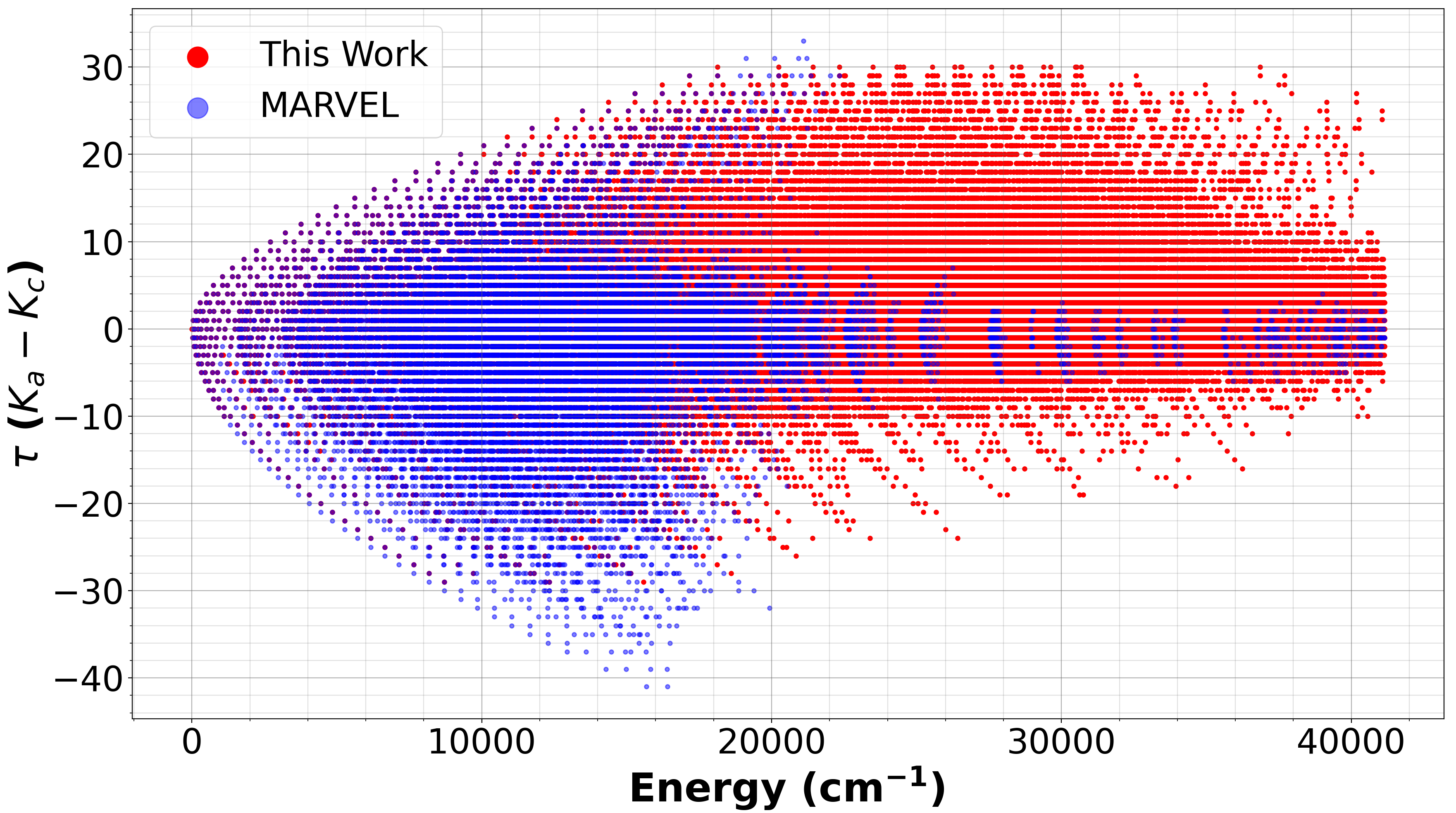}
	\caption{Our calculated states with $\sum_{\alpha\beta\gamma} (\Psi ^{J,k}_{\alpha\beta\gamma})^{2} >0.5$ and MARVEL \cite{jt817} states superimposed.}
	\label{fig:2}
\end{figure}

In Figure~\ref{fig:2} the total number of MARVEL states are plotted over the 68 610 states that we could uniquely label. Our predicted quantum labels directly overlap those values from MARVEL, proving our values are reliable and accurate. It is interesting to note the substantial amount of extra coverage in the regions where MARVEL lacks data, mostly in the visible (beyond approximately 20 000 cm$^{-1}$) and with $\tau=(K_{a}-K_{c})>0$. We cannot explain why the majority of assigned states appear to have $\tau=(K_{a}-K_{c})>0$, all we can assume at this stage is that $k$ becomes less well defined for $\tau=(K_{a}-K_{c})<<0$, i.e large values of $K_{c}$ corresponding to a highly oblate water molecule.

\begin{figure}
    \begin{subfigure}[]{0.5\textwidth}
    \centering
	\includegraphics[width=1.0\linewidth]{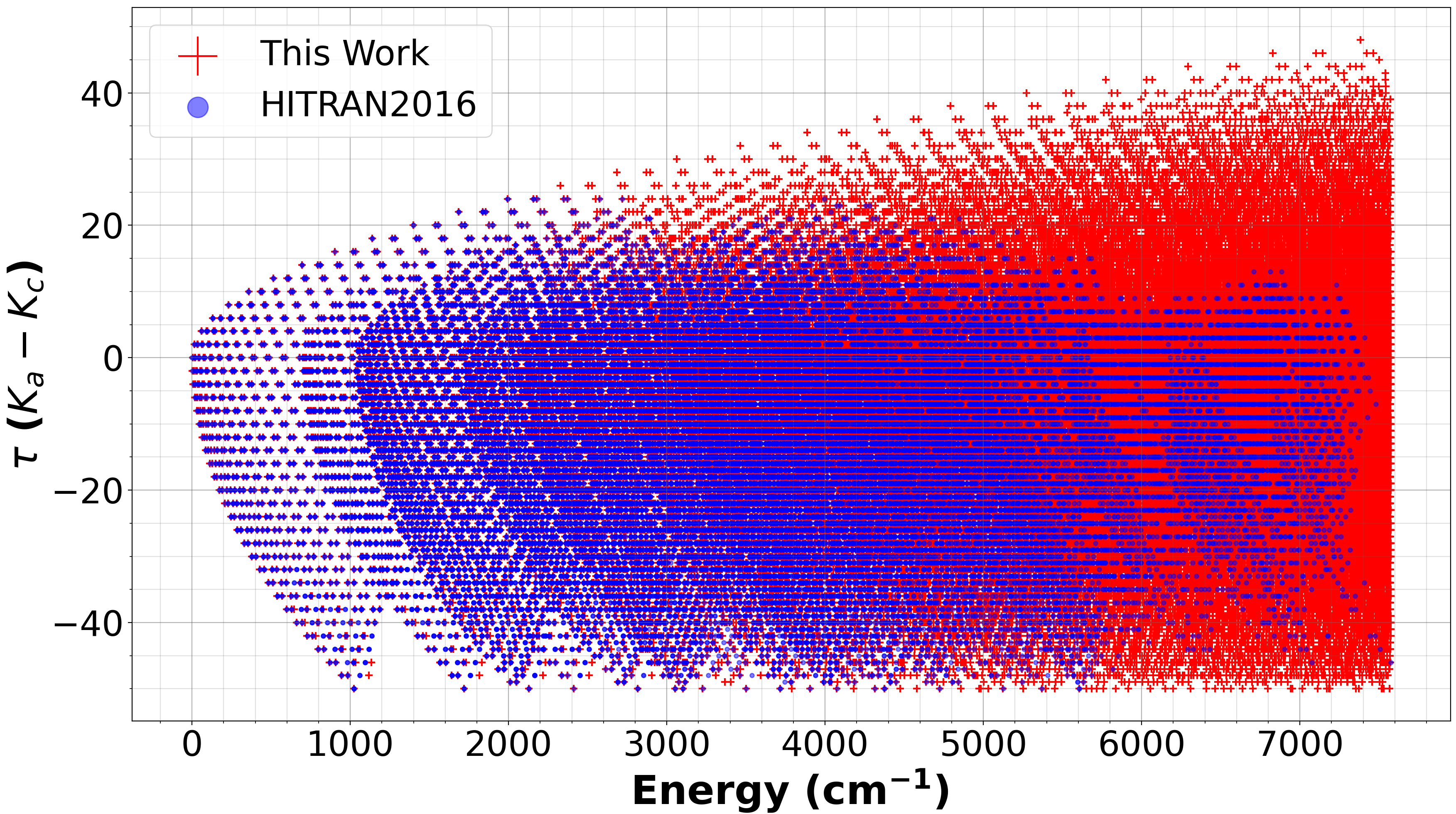}
	\caption{}
    \end{subfigure}
~
    \begin{subfigure}[]{0.5\textwidth}
    \centering
    \includegraphics[width=1.0\linewidth]{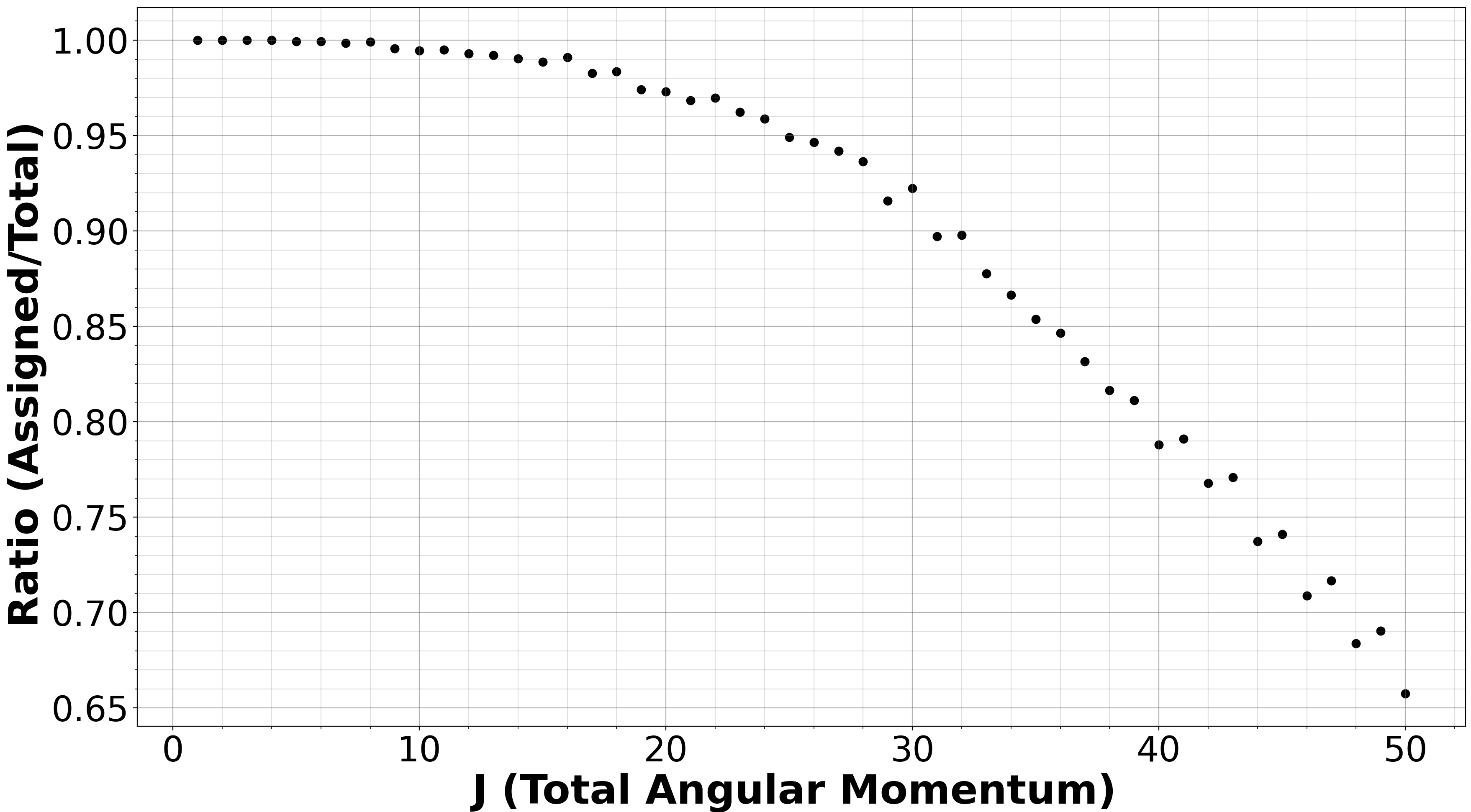}
    \caption{}
    \end{subfigure}
    
\caption{(a) Calculated O$_{3}$ states with $\sum_{\alpha\beta\gamma} (\Psi ^{J,k}_{\alpha\beta\gamma})^{2} >0.5$ and HITRAN2016 \cite{jt691} states superimposed. (b) The ratio of assigned states to the total number of available states (per $J$) as a function of increasing $J$.}
\label{fig:3}
\end{figure}

Ozone is a significantly more rigid molecule than water. Of the 192 812 calculated states, 169 840 of these have a value of $k$ that satisfies Eqn.~\ref{eqn:3}, which represents a 88\% success rate. As of yet, there has not been a MARVEL release on ozone, so we instead compare to the states available in the ozone line list within the HITRAN2016 \cite{jt691s} database that have values of total angular momentum less than or equal to 50 (note the HITRAN2016 line list extends to J=87). It should be noted that every state in the HITRAN2016 line list has been assigned vibrational and rotational quanta. In Figure~\ref{fig:3}(a), the 169 840 calculated states are plotted with those states in the HITRAN2016 line list. Our predicted quantum labels match the HITRAN2016 labels with a high degree of accuracy, evident from the overlap present in Figure~\ref{fig:3}(a). Unlike water, the number of states satisfying the Hose-Taylor theorem do not reduce in the limit of $K_{c}>>K_{a}$. However, for much higher values of J, this may indeed occur. 

In Figure~\ref{fig:3}(b), the ratio of the number of assignable states to the total number of states calculated per J is plotted for ozone. For $J=50$, over 65\% of calculated states are assignable, while for the water molecule, at $J=50$ less than 1\% of states were assignable. The noise present in Figure~\ref{fig:3}(b) is likely a consequence of us truncating the calculated states at the dissociation limit, which we took to be the upper limit of energies in the HITRAN2016 line list.

The DVR3D program suite has been updated and the code available on the ExoMol GitHub repository. Upon running the \textit{rotlev3b.f90} code, used to calculate wavefunctions, aside from writing the wavefunction files, it will simultaneously write out the values of $K_{a}$ and $K_{c}$ that satisfy the Hose-Taylor theorem together with the square of the wavefunction for that value of $k$.

\section{Conclusion}

We have demonstrated how the Hose-Taylor theorem  \cite{83HoTaxx.H2O} can be used with variational calculations to assign the rotational labels $K_{a}$ and $K_{c}$ for asymmetric top molecules using  water and ozone as examples.  The DVR3D program suite was used to calculate energy levels in a discrete variable representation in a Radau coordinate system. In principle, the theory should be applicable to any molecule regardless of its symmetry. Setting the maximum value of the total angular momentum for water and ozone molecules to 30 and 50 respectively, the Hose-Taylor theorem was used to assign $K_{a}$ and $K_{c}$ labels to almost 70 000 and 170 000 states for water and ozone respectively. The results indicate that for heavier, more rigid molecules, the quantity of assignable states increases. The Hose-Taylor theorem can be used to significantly improve our knowledge of local quantum labels for many molecules and ultimately increase the quality of spectroscopic information available. 

\section*{Acknowledgement}
Funding from the NASA AURA grant NNX17AI78G is acknowledged. This work was supported by  the UK Natural Environment
Research Council under grants NE/N001508/1 and the European Research Council under ERC Advanced Investigator grant 8838302. 


\end{document}